\documentclass{pasj00}
\SetRunningHead{M. Nobukawa et al.}
{3~K-shell lines of iron and 593-sec pulse from SAX~J1748.2$-$2808}
\Received{2008/05/19}
\Accepted{2008/06/22}
\begin{document}
\title{Discoveries of 3 K-shell Lines of Iron and a Coherent Pulsation 
of 593-sec from SAX J1748.2$-$2808}
\author{Masayoshi \textsc{Nobukawa}, Katsuji \textsc{Koyama},  
Hironori \textsc{Matsumoto}, and Takeshi Go \textsc{Tsuru}}
\affil{Department of Physics, Graduate school of Science, Kyoto 
University, Sakyo-ku, Kyoto 606-8502}
\email{nobukawa@cr.scphys.kyoto-u.ac.jp, koyama@cr.scphys.kyoto-u.ac.jp}
%%(Draft by Koyama and Data analysis by Nobukawa)
\maketitle

\begin{abstract}

SAX J1748.2$-$2808 is a unique X-ray object with
a flat spectrum and strong emission lines at 6.4--7.0~keV.
The Suzaku satellite resolved the emission lines into 
3 K-shell lines from neutral and highly ionized irons. 
A clear coherent pulsation with a period of 593-sec was
found from the Suzaku and XMM-Newton archives.
These facts favor that SAX J1748.2$-$2808 is
an intermediate polar, a subclass of magnetized white dwarf 
binary (cataclysmic variable: CV).
This paper reports on details of the findings and discusses the origin 
of this source.

\end{abstract}

\section{Introduction}

SAX J1748.2$-$2808 was discovered with the Beppo-SAX satellite 
in the direction of the radio complex Sagittarius (Sgr)~D \citep{Si01},
%%the giant molecular cloud Sgr~D (Sidoli et al. 2001),
and was deeply re-observed with the XMM-Newton satellite (Sidoli et al. 2006).
The X-ray spectrum was well-fitted with a power-law of photon index 1.4 plus 
a broad line of 0.43 keV ($1\sigma$) with the  center energy at 6.6 keV. 
The $N_{\rm H}$ value was as large as $1.4\times10^{23}$~H~cm$^{-2}$ 
(Sidoli et al. 2006). 
SAX J1748.2$-$2808 was thus regarded as one of the brightest samples of resolved 
point sources in the Galactic center region (GC) with Chandra deep-exposure 
observations (\cite{Mu03}, \cite{Mu06}). The integrated spectra of the 
point sources resemble the Galactic center diffuse X-rays (GCDX) in the 
iron line features close to 6.4--7.0 keV (\cite{Mu04}, \cite{Ko07b}).
Therefore, these, including SAX J1748.2$-$2808, can be regarded as 
significantly contributing to the GCDX, even though SAX J1748.2$-$2808 
was brighter than most of the other point sources in the GC. 

The XMM-Newton spectrum favored that SAX J1748.2$-$2808 is 
a High Mass X-ray Binary (HMXB) pulsar located at the GC region 
(Sidoli et al. 2006). However, no coherent pulsation was reported 
from either the Beppo-SAX or the XMM-Newton observations. 
Furthermore, the broad line of 0.43 keV ($1\sigma$) is very unusual as a HMXB. 
If this broadening is due to the Doppler effect, the velocity dispersion is 
as large as $2\times10^4$~km~s$^{-1}$. No such HMXB has been reported so far
(e.g. Nagase 1989). The detailed spectrum of  SAX J1748.2$-$2808 did not 
indicate whether the spectral nature is thermal or non-thermal.

In order to understand the nature of this unique object, we analyzed 
the Suzaku data of two pointing observations on the supernova remnants (SNR), 
Sgr D SNR and G~0.9+0.1, and also re-analyzed the XMM-Newton archive data.  

\section{Observations and Data Reduction}
The Suzaku and XMM-Newton observations of SAX J1748.2$-$2808 are 
listed in table~\ref{tab:obs_data}.
\begin{table*}
  \caption{Observation data list}
  \label{tab:obs_data}
  \begin{center}
    \begin{tabular}{lcccc}
      \hline\hline
      Observatory/Instrument & Target & Obs. ID & Date & Good Exposure Time* \\
      & & & (yyyy/mm/dd) & (ksec) \\
      \hline
      Suzaku/XIS & Sgr D SNR & 502020010 & 2007/09/06 & 139.1 \\
      Suzaku/XIS & G~0.9+0.1 & 502051010 & 2008/03/11 & 138.8 \\
      XMM-Newton/MOS & Sgr D SNR & 0112970101 & 2000/09/23 & 15.7  \\
      XMM-Newton/PN  & Sgr D SNR & 0112970101 & 2000/09/23 & 11.5  \\
      XMM-Newton/MOS & G~0.9+0.1 & 0144220101 & 2003/03/12 & 49.5 \\
      XMM-Newton/MOS & SAX J1748.2$-$2808 & 0205240101 & 2005/02/26 & 50.2 \\
      XMM-Newton/PN  & SAX J1748.2$-$2808 & 0205240101 & 2005/02/26 & 41.5 \\    
      \hline
      \multicolumn{5}{c}{* After the data screening described in the text.}
    \end{tabular}
  \end{center}
\end{table*}
The Suzaku satellite observed the two SNRs, Sgr D SNR and G 0.9+0.1, where 
SAX J1748.2$-$2808 was located near the edge of the field of view (FOV) 
of the XIS.
The XIS consists of four sets of X-ray CCD camera systems (XIS~0, 1, 2, and 3) 
placed on the focal planes of four X-Ray Telescopes (XRT) aboard the 
Suzaku satellite. XIS~0, 2, and 3 have front-illuminated (FI) CCDs, 
while XIS~1 has a back-illuminated (BI) CCD. One of the FI CCD cameras (XIS~2) 
has been out of function since September, 2006, and hence we did not use it.
Detailed descriptions of the Suzaku satellite, the XRT, and the XIS can 
be found in \citet{Mi07}, \citet{Se07}, and \citet{Ko07a}.

The XIS observations were made with the normal mode. The XIS pulse-height 
data for each X-ray event were converted to Pulse Invariant (PI) channels 
using the {\tt xispi} software version 2007-03, and the calibration database 
version 2008-01-31. We removed the data during the epoch of low-Earth 
elevation angles less than 5 degrees 
(ELV$<5^{\circ}$), day Earth elevation angles less than 10 degrees 
(DYE{\_}ELV$<10^{\circ}$), and the South Atlantic Anomaly.
The good exposure times are listed in table 1. Although the XIS CCDs 
were significantly degraded by on-orbit particle radiation,
the CCD performances were restored by the Spaced-low charge injection 
technique \citep{Uchi08}. 
Then, the overall spectral resolutions at 5.9 keV were $\sim$150 and
$\sim$170 eV (FWHM) for the FI and BI CCDs, respectively.
We analyzed the data using the software package HEASoft 6.4.1. 
For the spectral fittings, we made XIS response files using  {\tt xisrmfgen}, 
and auxiliary files using {\tt xissimarfgen}.
Since the spectrum of the non-X-ray background (NXB) 
depends on the geomagnetic cut-off rigidity (COR) \citep{Tawa2008NXB},
we obtained COR-sorted NXB spectra using {\tt xisnxbgen}, from the night-Earth data 
released by the Suzaku XIS team. 

SAX J1748.2$-$2808 was also in the filed of view of the XMM-Newton observations
on Sgr D SNR and G~0.9+0.1 on September 2000 and March 2003, respectively. 
The pointing observation on SAX J1748.2$-$2808 was also made on February, 2005.
The X-ray data were obtained with the European Photon Imaging Camera (EPIC) 
(\cite{St01}; \cite{Tu01}) in 
an extended full-frame mode. The data were analyzed using the Science 
Analysis Software (SAS~7.1.0). Event files for both the PN and the Metal 
Oxide Semiconductor (MOS) detectors were produced using the {\tt epchain} and 
{\tt emchain} tasks of SAS, respectively. 
The event files were screened for high particle-background periods. 
Good exposure times are listed in table 1. In our analysis, we used 
the events corresponding to patterns 0--4 for the PN and 0--12 for 
the MOS instruments. The PN data of SAX J1748.2$-$2808 on the 2003 
observation was not available.

\section{Analysis and Results} % section 3

\subsection{X-ray Image} %3-1

The X-ray images taken with the XIS were analyzed for each observation separately.
Since the Suzaku nominal position error is $\lesssim \timeform{20"}$ \citep{U08}, 
we fine-tuned 
the Suzaku position while referring to the positions of XMM-Newton sources. 
In the Sgr D SNR observation, we found an XMM-Newton source S10 (see table 4 
of Sidoli et al. 2006). The XMM-Newton position of S10 is 
($\alpha,\ \delta$) = ($\timeform{267D.24531},\ \timeform{-28D.188918}$), 
while the\ Suzaku position is 
($\alpha,\ \delta$) = ($\timeform{267D.24409},\ \timeform{-28D.185869}$).
We therefore shifted the Suzaku coordinate by 
($\Delta\alpha,\ \Delta\delta$) = ($\timeform{0D.00122},\ \timeform{-0D.003049}$).
In the G0.9+0.1 observation, we used an XMM source S2 (see table 4 of \cite{Si06}).
The XMM Newton position is 
($\alpha,\ \delta$) = ($\timeform{266D.81562},\ \timeform{-28D.181511}$), 
while the Suzaku position is 
($\alpha,\ \delta$) = ($\timeform{266D.81737},\ \timeform{-28D.180065}$). 
We hence shifted the Suzaku coordinate by ($\Delta\alpha,\ \Delta\delta$) = 
($\timeform{-0D.00175},\ \timeform{-0D.001446}$).
\begin{figure*}[htpb] %figure 1
\begin{center}
\FigureFile(160mm,80mm){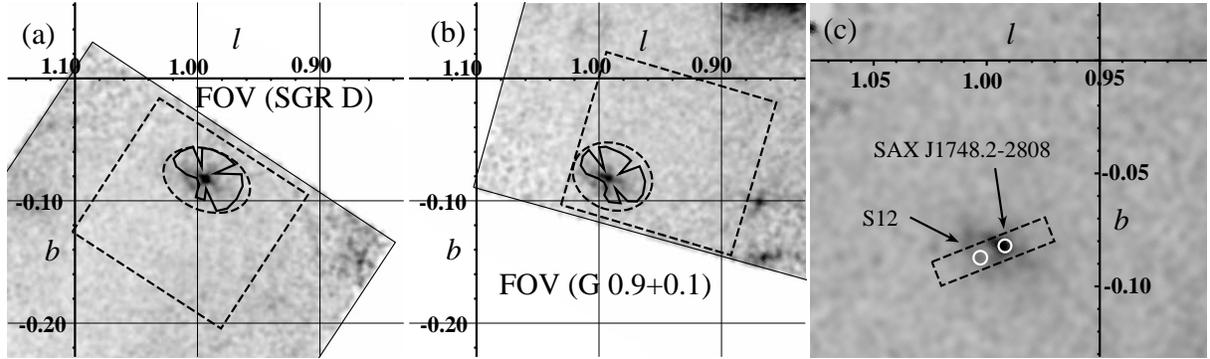}
\end{center}
\caption{X-ray image in the 3--7 keV band from the Suzaku observations on 
Sgr D SNR (a) and G 0.9+0.1 (b). 
The source regions of SAX J1748.2$-$2808 are given by the solid polygons, 
which trace the complicated point-spread function of the XRT in the field edge 
(see text). The dashed squares (excluding ellipses) are the local backgrounds. 
(c): Combined images of figures (a) and (b) near SAX J1748.2$-$2808. 
The projected profile (figure 2) was made from the dashed rectangle region, 
where the white circles indicate the positions of SAX J1748.2$-$2808 
(stronger source) and S12 (fainter source) \citep{Si06}.} 
\label{fig:Map}
\end{figure*}
After these fine-tunings of the Suzaku coordinate, we made the X-ray image 
shown in figure~1. We found two sources near to the edge of each field. 
These two sources coincide in position with S3, i.e., SAX J1748.2$-$2808 
(stronger source) and S12 (fainter source) of the XMM-Newton observation 
within statistical errors of $\lesssim \timeform{5"}$. 

In order to estimate the intensity ratio of the two sources, we made a projected 
profile along the line connecting SAX J1748.2$-$2808 and S12 (see figure~1c).  
The profile was fitted with two Gaussians plus a liner function, as is shown 
in figure~2. The widths (1-$\sigma$) of the Gaussians approximate the projected 
point-spread function. The best-fit source fluxes (normalizations of the Gaussians) 
were determined to be $7.5\times10^{-3}$ and $2.8\times10^{-3}$ counts s$^{-1}$, 
for SAX J1748.2$-$2808 and for S12, respectively. 

%%\begin{figure}[h] %figure 2
%%\begin{center}
%%\FigureFile(80mm,50mm){figure2.eps}
%%\end{center}
%%\caption{Projected profile along the line connecting SAX J1748.2$-$2808 
%%and S12. This profile was made from the data in the dashed rectangle in 
%%figure~1c. The solid lines are the best-fit liner function and 2-Gaussians.}
%%\label{fig:Profile}
%%\end{figure}

\subsection {Spectrum} %3-2

Since the X-ray spectrum of XMM-Newton has already been reported, we show 
the Suzaku spectrum of SAX~J1748.2$-$2808. The point-spread function of the 
XRT has a complicated polygon shape images, due together to the 4-segments 
structure of the XRT and image deformation near the field edge (\cite{Se07}). 
We therefore extracted the spectra from the solid polygons in figures 1a and 1b, 
for the best S/N ratio within limited statistics. 

The background spectra were obtained from a near-by sky shown by the 
dashed squares, from which the dashed elliptical regions were excluded. 
The background spectra consist of the non-X-ray background (NXB) and 
the local background (the cosmic X-ray background plus the Galactic center 
diffuse X-rays). Since the local background is affected by the 
vignetting of the XRT while the NXB is not, each background was
treated separately.
For both the source and the background data, we made COR distributions and 
composed the NXB spectra from the COR-sorted NXB data set (see section 2).
We then subtracted the NXB extracted from the same detector areas. 
After NXB subtraction, 
we corrected the vignetting effect due to the different off-axis 
angles between the source and the background regions by multiplying the 
effective-area ratios for each energy bin of the local background spectra
(the same process as Hyodo et al. 2008).
We further subtracted the corrected local background from the source spectra. 
All of the source spectra of the two FIs from the two observations are 
co-added to increase the statistics.

We estimated the contamination of the near-by fainter source S12 
by a ray-tracing simulation ({\tt xissim} in the HEASoft package).
The spectral shape of SAX J1748.2$-$2808 was assumed to be an absorbed 
power-law plus 3 Gaussian lines, whose parameters are the same as table~2. 
This spectral model will be mentioned in the next paragraph.
Since the fainter source S12 is too weak to allow a spectral analysis, 
its spectral shape was assumed to be the same model of an absorbed black-body 
($N_{\rm H}=10^{24}$~cm$^{-2}$, k$T=$1~keV) as \citet{Si06}. 
Using the flux ratio of SAX~J1748.2$-$2808 and S12 determined in section~3.1,
we found that the contamination was $\lesssim$10\% in the 3--10 keV band. 
We therefore ignore the contamination of S12 in the following analysis 
and discussion. 

The X-ray spectra of FI and BI were simultaneously fitted with an 
absorbed power-law plus one broad line, Abs$\times$(Power-Law + 1 Gaussian line), 
which is the same model as that of the XMM-Newton by \citet{Si06}. 
The best-fit (90\% confidence range) photon index of the power-law, 
line center energy, and width ($1\sigma$) are 
1.2 (0.8--1.5), 6.66 (6.54--6.77)~keV, and 0.31 (0.21--0.39)~keV, 
respectively ($\chi^2$/dof of 72.8/64).
These parameter values are consistent with those of XMM-Newton 
within the statistical error. However, the Suzaku spectrum exhibits 
significant residuals near to energies of 6.4, 6.7, and 7.0 keV. 
We therefore fitted with a model of an absorbed power-law plus 3 narrow 
lines near to 6.40, 6.68 and 6.97~keV, Abs$\times$(Power-Law + 3 Gaussian 
lines). Although the fittings were made simultaneously for the FI and BI 
spectra, we simply show the FI result in figure 3. The fit was improved 
with $\chi^2$/dof of 67.6/64. The best-fit parameters are listed in table 2. 

\begin{figure}[h] %figure 2
\begin{center}
\FigureFile(80mm,50mm){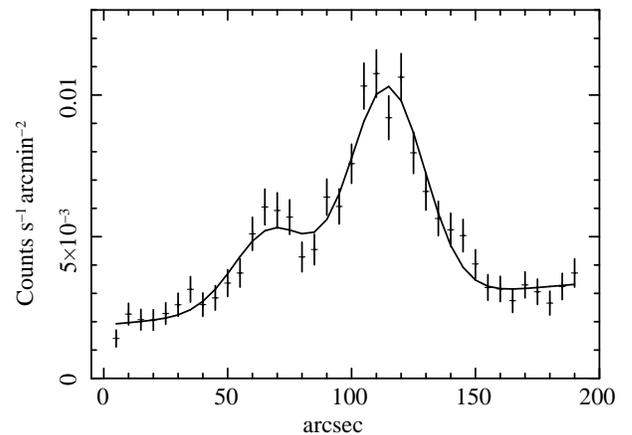}
\end{center}
\caption{Projected profile along the line connecting SAX J1748.2$-$2808 
and S12. This profile was made from the data in the dashed rectangle in 
figure~1c. The solid lines are the best-fit liner function and 2-Gaussians.}
\label{fig:Profile}
\end{figure}

\begin{figure*} %figure 3
\begin{center}
\FigureFile(120mm,80mm){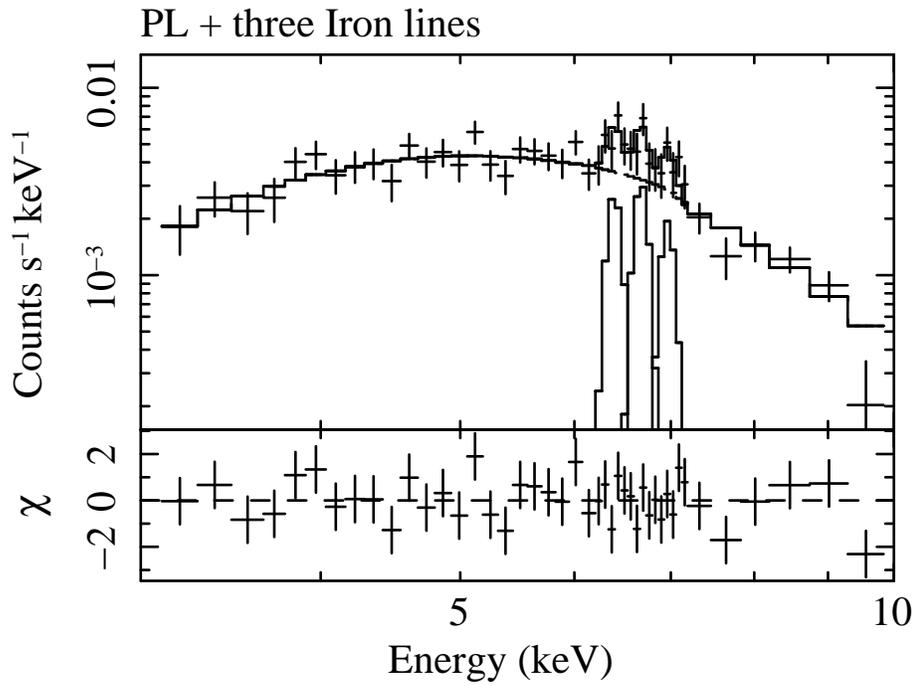}
\end{center}
\caption{Background-subtracted spectrum of SAX~J1748.2$-$2808 for the FI CCD.
The source spectrum was made from the solid polygons in figures 1a and 1b, while
the local background spectra were taken from the dashed squares excluding the 
dashed ellipses.
The solid line is the best-fit model of an absorbed power-law plus 3 Gaussians.}
\label{fig:Spec3}
\end{figure*}

The best-fit line energies of 6.7 keV and 7.0 keV are likely due to K$\alpha$ 
lines from He-like and H-like irons at energies of 6.68keV and 6.97 keV, 
respectively, while the 6.4 keV line would be a K$\alpha$ line from neutral irons.
The 6.97 keV line may be contaminated by a K$\beta$ line (7.05 keV) of neutral irons. 
We therefore applied a more physical model: an absorbed thin thermal plasma (APEC) 
plus the 6.40 keV and 7.05 keV lines from neutral irons, 
Abs$\times$(APEC + 6.40~keV + 7.05~keV lines). In this model, we fixed 
the flux ratio of the 6.40 keV and 7.05 keV lines to 1: 0.125 
(Kaastra \& Mewe 1993). A simultaneous fit for the FI and BI spectra with 
$\chi^2$/dof of 77.4/65 is acceptable at the 90\% confidence level,
as is given in table 2. In figure~4, we show only the FI spectrum for simplicity.

\begin{figure*} %figure 4
\begin{center}
\FigureFile(120mm,80mm){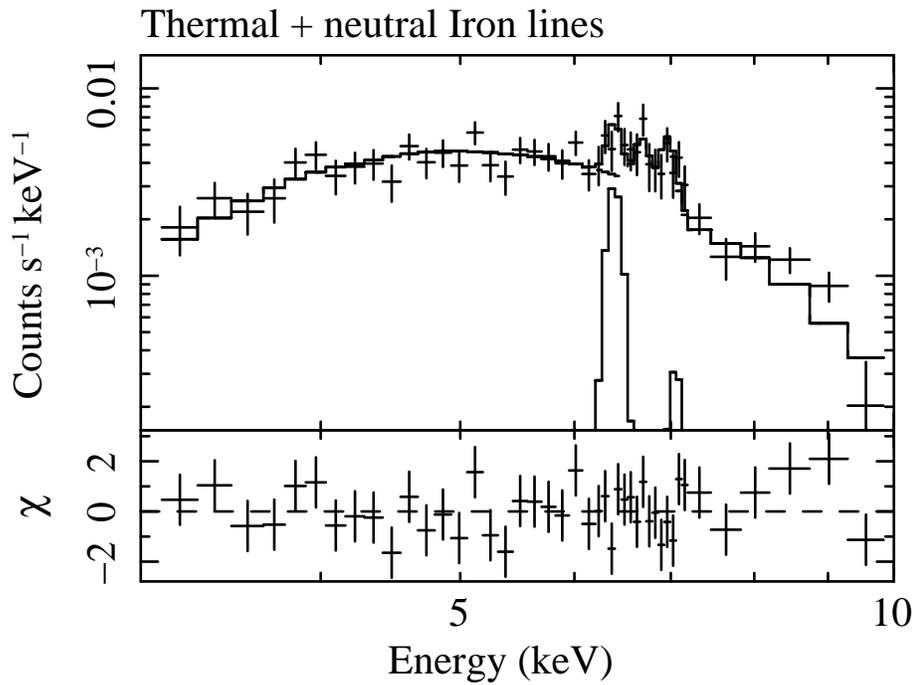}
\end{center}
\caption{Same as figure 3, but the best-fit model of an absorbed thin thermal 
plasma (APEC) plus K-shell lines of neutral iron.}
\label{fig:Spec4}
\end{figure*}

\begin{table*} %table 2
  \caption{Best-fit parameters}
  \label{tab:spectrum}
  \begin{center}
    \begin{tabular}{lll}
      \hline\hline
\multicolumn{3}{c}{Model: Abs$\times$(Power-Law + 3 Gaussian  Lines)}\\
\hline
Line Energy (keV) & Flux (10$^{-6}$ ph s$^{-1}$ cm$^{-2}$) & Equivalent Width (eV)\\
\hline
6.40~(6.39$-$6.47) & 1.5~(0.8$-$2.3) &140~(30$-$270) \\
6.68~(6.66$-$6.72) & 1.9~(1.1$-$2.7) &180~(30$-$350) \\
6.97~(6.94$-$7.46) & 1.3~(0.6$-$2.1) &130~($\leq 270$) \\
\hline
Parameter & Bets-Fit Value\\
\hline
Photon Index & 1.0~(0.8$-$1.3) \\
$N_{\rm H}$~($10^{23}$~cm$^{-2}$) & 1.3~(1.0$-$1.7)\\
Flux (3$-$10~keV)~($10^{-13}$~erg~s$^{-1}$~cm$^{-2}$) & 6.0~(5.7$-$6.2)\\
Luminosity$^\dag$ (3$-$10~keV)~($10^{33}$~erg~s$^{-1}$) & 7.4~(7.0$-$7.6)\\
$\chi^2$ / dof & 67.6 / 64 & \\
\hline\hline
\multicolumn{3}{c}{Model: Abs$\times$(APEC + Neutral Iron Lines)}\\
\hline
Line Energy (keV)  & Flux (10$^{-6}$ ph s$^{-1}$ cm$^{-2}$) & Equivalent Width (eV)\\
\hline
6.40 (fixed) & 2.1~(1.0$-$2.7) & 160~(10$-$270)\\
7.05 (fixed) & 0.26$^\ddag$ & 23 \\
\hline
Parameter & Best-Fit Value\\ 
\hline
Temperature: k$T$ (keV) & 12~(9$-$17)\\
Metal abundance (solar) & 0.57 (0.35-0.85) \\
$N_{\rm H}$~($10^{23}$~cm$^{-2}$) & 1.8~(1.6$-$2.1)\\
Flux (3$-$10~keV)~($10^{-13}$~erg~s$^{-1}$~cm$^{-2}$) & 5.4~(5.2$-$5.7)\\
Luminosity$^\dag$ (3$-$10~keV) ($10^{33}$~erg~s$^{-1}$) & 8.4~(8.0$-$8.8)\\
$\chi^2$ / dof & 77.4 / 65\\
\hline\hline
\multicolumn{3}{l}{* Parenthesis is 90\% confidence range.}\\%%[-5pt]
\multicolumn{3}{l}{$^\dag$ Absorption corrected. 
Distance toward SAX J1748.2$-$2808 is assumed to be 8.5~kpc.}\\%%[-5pt]
\multicolumn{3}{l}{$^\ddag$ The flux of the 7.05 keV line is constrained to be 
0.125 of the 6.40 keV Line.}\\
\end{tabular}
\end{center}
\end{table*}

\subsection{Timing}

We examined a long-term (8 years) X-ray flux history from the 5 observations 
listed in table 1. The X-ray fluxes were calculated by fitting the 3-line model 
with the parameters fixed values given in table~2. Only 
the normalization (flux) was a free parameter. The resultant fluxes in the 
3$-$10 keV band show no significant variation during the 8 years interval. 

We then searched for a coherent pulsation in the 3$-$7~keV band from all of 
the observations listed 
in table~1. The Fast Fourier Transform (FFT) analysis revealed a clear peak 
at $\sim1.7\times10^{-3}$ Hz from all of the observations.
We then searched for accurate pulse period with the folding technique, and
found a significant peak near to the trial period of 593 sec. 
The best-fit pulse periods and errors are listed in table 3 for all of the 
observations. As an example, we show the Suzaku results from 2007: the power 
spectrum of FFT, periodogram, and folded light curve in figure 5a, 5b, 
and 5c, respectively.  

The 593-sec pulsation is likely to be a spin rotation of either a magnetic white dwarf 
or a neutron star in a binary system (see section 4). We, therefore, searched for 
an orbital modulation in the light curve with 1000-sec binning. 
In the Suzaku observations, we found no sign of orbital modulation, nor any sign of 
an eclipse. 
And no orbital modulation was found from the XMM-Newton data, either.
This is reasonable because the time coverages of the XMM-Newton observations 
were less than the Suzaku.

\begin{figure*} %figure 5
\begin{center}
\FigureFile(80mm,50mm){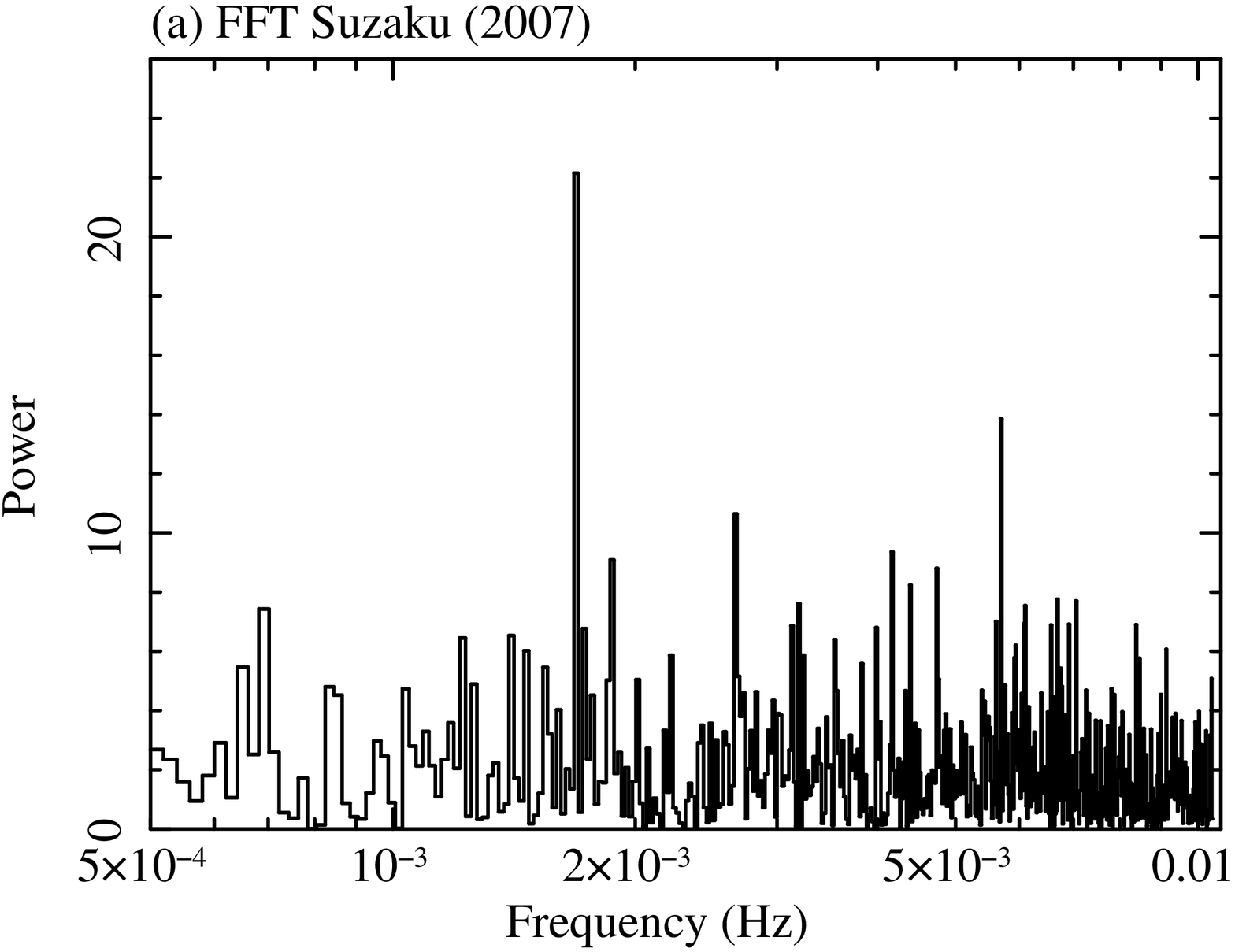}
\FigureFile(80mm,50mm){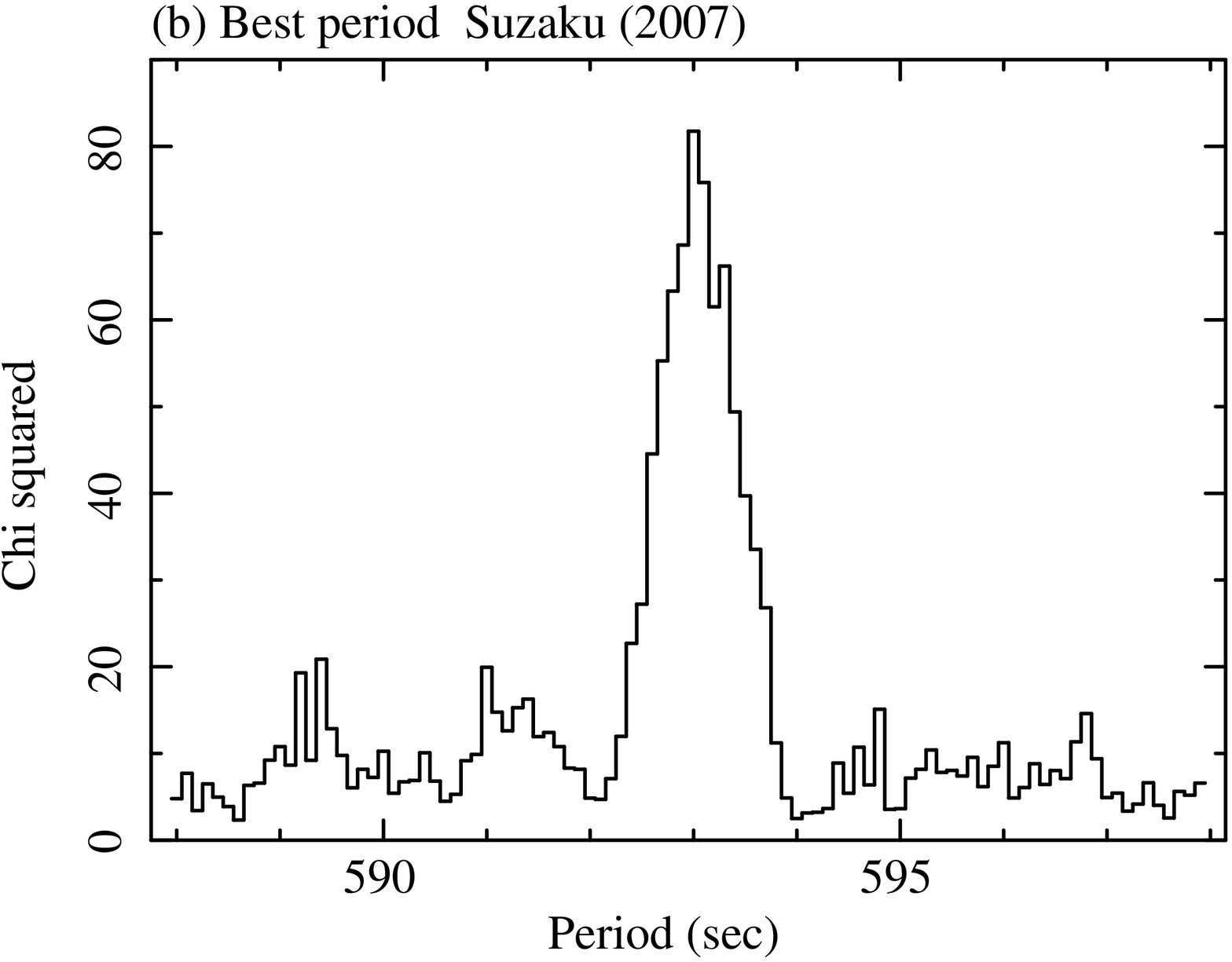}
\FigureFile(80mm,50mm){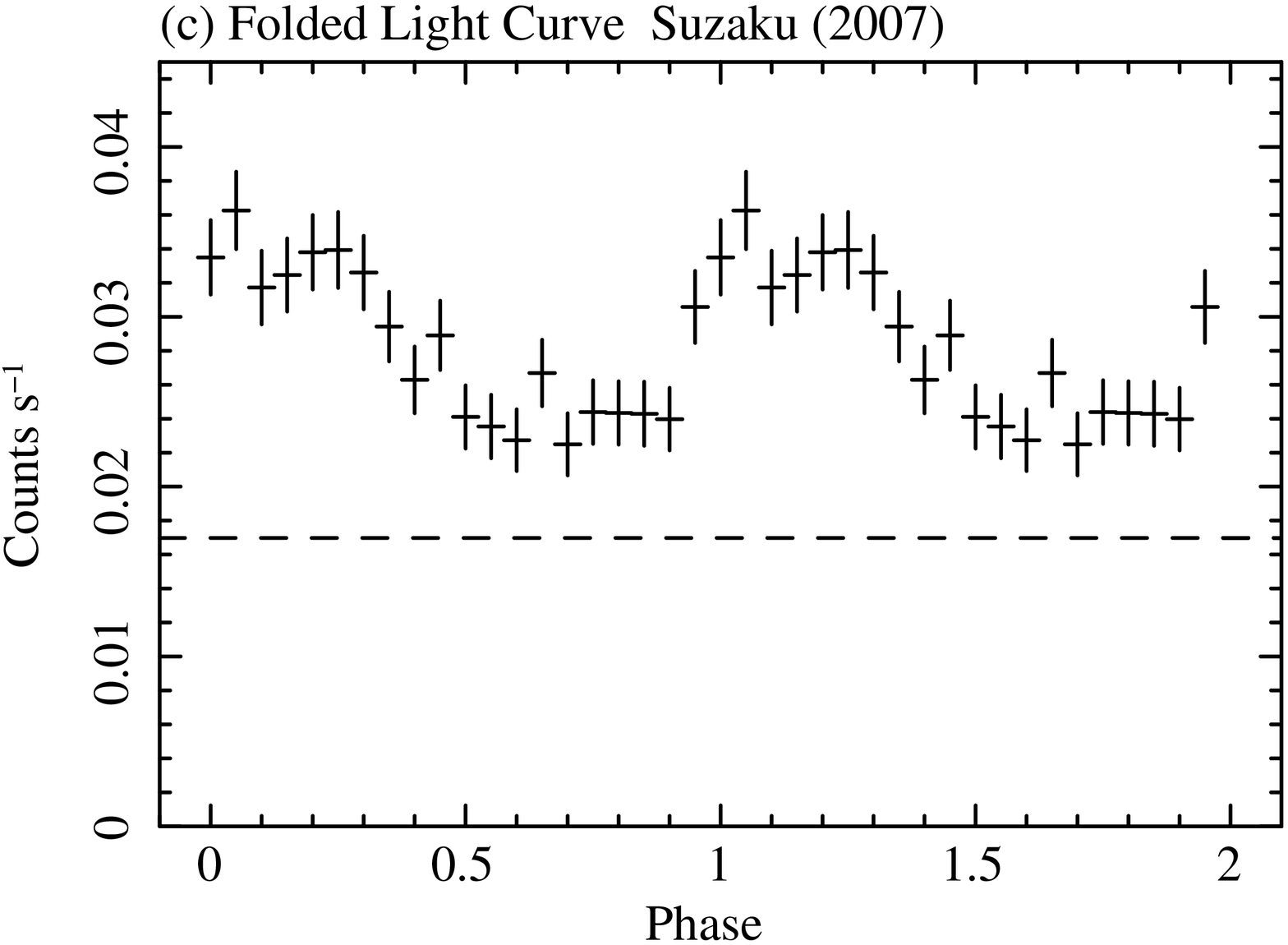}
\end{center}
\caption{(a): Power spectrum (FFT) in the 3$-$7~keV band of the Suzaku 
observation (2007).
(b): Same as (a), but the periodogram at around the trial period of 593 sec.
(c): Same as (a), but the folded light curve at the best-fit period of 593.1~sec,
 where the dashed line is the background level.}
\label{fig:Time5}
\end{figure*}

\begin{table*} %table 3
  \caption{Best-fit Period}
  \label{tab:timing}
  \begin{center}
    \begin{tabular}{lll}
      \hline\hline
      Observatory (Year/Month) & Instruments &Pulse Period (s)\\
      \hline
      XMM-Newton (2000/09) & MOS1+2 & $592 \pm8$ \\
      XMM-Newton (2000/09) & PN & $594 \pm8$ \\
      XMM-Newton (2003/03) & MOS1+2 & $593 \pm2$\\ 
      XMM-Newton (2005/02) & MOS1+2 & $595 \pm3$ \\
      XMM-Newton (2005/02) & PN & $593 \pm2$ \\
      Suzaku (2007/09) & XIS~0+1+3 & $593.1 \pm0.4$ \\
      Suzaku (2008/03) & XIS~0+1+3 & $592.8 \pm0.4$ \\     
      \hline
      \multicolumn{3}{l}{* Error is 1-$\sigma$ of the Gaussian of the folded result 
        (e.g. figure~\ref{fig:Time5}b).} \\
    \end{tabular}
  \end{center}
\end{table*}

\section {Discussion}
 
We found a coherent pulsation of 593-sec from SAX J1748.2$-$2808. This
constrains the origin of this source to be either a magnetic CV, or a HMXB pulsar.
The 593-sec period is among the slowest 10\% of the pulse period in HMXB pulsars 
(\cite{Na89}, \cite{Li00}), while the fastest 30\% in magnetic CVs (\cite{Ri03}). 
Thus, the spin period of 593-sec favors a magnetic CV scenario, although a HMXB 
pulsar scenario is not firmly excluded.

The Suzaku spectrum resolved the broad line at 6.6 keV found in the previous 
XMM-Newton observation into, at least,  three lines at 6.40, 6.68, 6.97 keV.
The best-fit equivalent width ($EW$) of these three lines are $\sim$140~eV, 
$\sim$180~eV, and $\sim$130~eV, respectively.  
Ezuka and Ishida (1999) compiled the ASCA data and reported
that the magnetic CVs exhibit 3 iron K$\alpha$ lines at 6.4, 6.7, and 7.0 keV with 
mean $EW$ values of $\sim$100, $\sim200$, and $\sim$100~eV, respectively 
(see also \cite{He98}), nearly the same as those of SAX J1748.2$-$2808.

The overall spectrum was also well fitted with a thin thermal plasma of 12-keV 
temperature with a sub-solar iron abundance plus K$\alpha$ (6.40~keV) and 
K$\beta$ (7.05~keV) lines from neutral irons. 
Ezuka and Ishida (1999) also reported that the spectra of the magnetic CVs
can be described by a thin thermal plasma model plus 6.4 keV line
with a mean temperature of $\sim$20 keV, closely resemble to the Suzaku 
results of SAX J1748.2$-$2808.
On the other hand, HMXB pulsars show a hard continuum spectrum, but it is 
a broken power-law and not 
a thin thermal. HMXBs exhibit an iron emission line feature. However, the line 
feature is not complex, but is a single line, mostly at 6.4~keV (see table 3 
of \cite{Na89}). These properties are different from those of SAX J1748.2$-$2808. 

All of the above facts favor the idea that SAX J1748.2$-$2808 is a magnetic CV  
rather than a HMXB. The pulse period of 593-sec is smaller than the possible 
orbital period, although we found no orbital modulation. Therefore, 593-sec 
would be the spin period, and hence SAX J1748.2$-$2808 is an intermediate polar 
(IP), not a polar with synchronized spin and an orbital period. 
If the luminosity of SAX J1748.2$-$2808 is typical value for an IP of 
$\sim 10^{33}$~erg~s$^{-1}$ \citep{Pa94}, then from the observed flux of $\sim 
6\times10^{-13}$~erg~cm$^{-2}$~s$^{-1}$, the distance is estimated to be $\sim$4~kpc.
Thus, SAX J1748.2$-$2808 would be a foreground source, not a member of the GC sources.

Conversely, \citet{Si06} suspected that SAX J1748.2$-$2808 is located near the GC 
region at about 8.5~kpc, because it has a large absorption of 
$1.4\times10^{23}$~H~cm$^{-2}$.
With this distance, they estimated the luminosity to be $\sim 10^{34}$~erg~s$^{-1}$, 
significantly larger than any other magnetic CVs, and hence the authors declined 
to suggest a HMXB origin. We, however, note that a large fraction of the absorption 
of IPs is due to the circum-stellar gas, and the absorption value could be 
up to several $\times 10^{23}$~H~cm$^{-2}$ \citep{Ez99}. Therefore, the large 
absorption of SAX J1748.2$-$2808 would also be due to circum-stellar gas, rather
than interstellar gas integrated along the long line of sight.
This large amount of circum-stellar gas can be naturally explained as being the 
origin of the strong 6.4 keV line from neutral irons (see e.g. \cite{Ez99}).

\bigskip

The authors thank all of the Suzaku team members, especially Y. Hyodo, 
H. Uchiyama, M. Ozawa, H. Nakajima, H. Yamaguchi, and H. Mori for their 
supports and useful information on the XIS performance.
We also thank our referee, M. Ishida for his useful comments.
This work is  supported by Grant-in-Aids from the Ministry of Education, 
Culture, Sports, Science and Technology (MEXT) of Japan,  Scientific Research A (KK), 
and  Grant-in-Aid for Young Scientists B (HM).
HM is also supported  by the Sumitomo Foundation, Grant for Basic Science 
Research Projects, 071251, 2007. MN is supported by JSPS Research Fellowship 
for Young Scientists.

\end{document}